\documentstyle[psfig]{mn}

\input{epsf}

\begin{document}

\title[The absolute infrared magnitudes of type~Ia supernovae]{The absolute 
infrared magnitudes of type~Ia supernovae}

\author[W.P.S. Meikle]
{ W.P.S. Meikle$^{1}$ \\
$^1$Astrophysics Group, Blackett Laboratory, Imperial College of
Science, Technology and Medicine, Prince Consort Road, \\ London 
SW7 2BW \\
}

\date{Accepted 2000 January 6.  Received 2000 January 6; in original form 1999 October 13}

\pagerange{\pageref{firstpage}--\pageref{lastpage}}
\pubyear{1994}

\maketitle

\label{firstpage}

\begin{abstract}
The absolute luminosities and homogeneity of early-time infrared (IR)
light curves of type~Ia supernovae are examined.  Eight supernovae are
considered.  These are selected to have accurately known epochs of
maximum blue light as well as having reliable distance estimates
and/or good light curve coverage.  Two approaches to extinction
correction are considered.  Owing to the low extinction in the IR, the
differences in the corrections via the two methods are small.
Absolute magnitude light curves in the $J$, $H$ and $K$-bands are
derived.  Six of the events, including five established
``Branch-normal'' supernovae show similar coeval magnitudes.  Two of
these, SNe~1989B and 1998bu, were observed near maximum infrared
light.  This occurs about 5~days {\it before} maximum blue light.
Absolute peak magnitudes of about --19.0, --18.7 and --18.8 in $J$,
$H$ \& $K$ respectively were obtained.  The two spectroscopically
peculiar supernovae in the sample, SNe~1986G and 1991T, also show
atypical IR behaviour.  The light curves of the six similar supernovae
can be represented fairly consistently with a single light curve in
each of the three bands.  In all three IR bands the dispersion in
absolute magnitude is about 0.15~mag, and this can be accounted for
within the uncertainties of the individual light curves.  No
significant variation of absolute IR magnitude with $B$-band light
curve decline rate, $\Delta m_{15}(B)$, is seen over the range
$0.87<\Delta m_{15}(B)<1.31$.  However, the data are insufficient to
allow us to decide whether or not the decline rate relation is weaker
in the IR than in the optical region.  IR light curves of type~Ia
supernovae should eventually provide cosmological distance estimates
which are of equal or even superior quality to those obtained in
optical studies.
\end{abstract}


\section{Introduction}
It is increasingly believed that the optical light curves of type~Ia
supernovae (SNe~Ia) differ between events in intrinsic peak luminosity
and shape.  Moreover, the SN~Ia peak luminosity appears to correlate
with its optical decline at early times; slower declining SNe~Ia have
greater peak luminosity (Pskovskii 1977, 1984; Phillips 1993; Hamuy
{\it et al.} 1995; Riess, Press \& Kirshner 1995; Hamuy {\it et al.}
1996a; Riess {\it et al.} 1998; Saha {\it et al.} 1999; Phillips {\it
et al.} 1999).  Work by Hamuy {\it et al.} (1996a), Tripp (1998),
Saha {\it et al.} (1999) and Phillips {\it et al.}  (1999) indicate
slopes of between 0.5 and 0.8 for the variation of the absolute $B$
magnitude at maximum light, $M_{Bmax}$, with $\Delta m_{15}(B)$, the
decline in $B$ magnitude at 15~days post-$t_{Bmax}$. ($t_{Bmax}$ is
the epoch of maximum light in the $B$-band.  Supernova epochs in this
paper will usually be relative to this fiducial time.) It is suggested
by some that the relation also contains a non-linear component (Riess
{\it et al.}  1998, Saha {\it et al.} 1999, Phillips {\it et al.}
1999). \\

It has been established that, even after correction has been made for
the decline-rate relations, the absolute magnitude at $t_{Bmax}$
varies with colour.  Phillips {\it et al.} (1999) endeavour to derive
decline-rate relations free of the effects of host galaxy dust
reddening.  They find that, in a sample which includes events which
are noteably red at maximum {\it viz.} $(Bmax-Vmax)>0.2$, a plot of
$M_{Bmax}$ against $(Bmax-Vmax)$ (both decline-rate corrected) yields
a slope of 3.5$\pm$0.4 which is consistent with the Galactic reddening
law.  Even if the sample is restricted to those SNe~Ia having
$(Bmax-Vmax)<0.2$ (the great majority), Phillips (private
communication) still finds a slope of $3.9\pm1.3$.  This tends to
support their contention that most of the decline-rate corrected
colour variation is due to dust reddening. \\

Tripp (1998) and Saha {\it et al.} (1999) examine samples restricted
to $(Bmax-Vmax)<0.2$. These are very similar to the $(Bmax-Vmax)<0.2$
subset of the Phillips {\it et al.} sample.  Unlike Phillips {\it et
al.}, neither Tripp nor Saha {\it et al.} try to separate out possible
competing effects of intrinsic colour variation and dust reddening in
these samples.  For their decline-rate corrected $M_{Bmax}$ versus
$B-V$ slope, Tripp (1998) finds $2.4\pm0.4$ and Saha {\it et al.}
(1999) find 1.7, smaller than the value found by Phillips {\it et al.}
(although, as indicated above, the uncertainties are large).  If
extinction is negligible in the $(Bmax-Vmax)<0.2$ SN samples then the
Tripp and Saha {\it et al.} results may indicate that significant
intrinsic colour variation in SNe~Ia is present even after
decline-rate correction.  However, if the colour variation has a
significant component due to dust extinction effects then it is less
clear what the results of Tripp and Saha {\it et al.} signify. \\

In order to exploit fully the SN~Ia phenomenon as a valuable distance
indicator, it is desirable to find as many ways as possible for
dealing with the effects of extinction.  In this paper I examine light
curves in the near-infrared ($JHK$) where it might be hoped that the
effects of interstellar dust are less.  This approach was anticipated
in the pioneering paper of Elias {\it et al.} (1981), where high
quality $JHK$ light curves were presented for the type~Ia supernovae
1980N, 1981B and 1981D.  Two of these, SNe~1980N and 1981D occurred in
the same galaxy, NGC~1316 (Fornax~A).  By simply shifting the SN~1981B
light curves by 0.4~mag, they found that the resulting scatter in the
individual light curves appeared to be 0.1~mag or less, which was
barely larger than the uncertainties in the photometry.  This was a
lower scatter than that seen in optical light curves at that time, and
Elias {\it et al.} suggested that this was due to the much reduced
effects of internal reddening in the IR.  They suggested that,
consequently, IR light curves of supernovae might be useful as
distance indicators out to a few tens of Mpc.  In Elias {\it et al.}
(1985), fiducial IR light curves (``templates'') were produced, made
up of a series of straight line segments.  These were based on $JHK$
data points for the three supernovae from Elias {\it et al.} (1981)
together with additional IR photometry from eight other supernovae.
The light curves of individual supernovae were adjusted in relative
magnitude to provide the best fits to the fiducial curves.  Relative
epochs were fixed by the estimated $t_{Bmax}$. In spite of the
sometimes large uncertainty in $t_{Bmax}$, they found that the
uncertainty in the fiducial curves was roughly $\pm$0.03~mag between
days t=+5 and t=+40. They suggested that SNe~Ia have a dispersion in
absolute IR magnitude of $\pm$0.2~mag, and possibly $\pm$0.1~mag,
making them potentially valuable for distance determination within the
Local Supercluster. \\

Before the work of Elias {\it et al.}, Kirshner {\it et al.} (1973)
acquired the first ever IR measurements of a SN~Ia {\it viz.}
SN~1972E, but only a part of their data are of quality comparable to
that of modern photometry.  Between the publications of Elias {\it et
al.} (1981, 1985) and the advent of SN~1998bu, very little IR
photometry of SNe~Ia was reported.  While good coverage of the
SN~1986G IR light curves was achieved by Frogel {\it et al.} (1987),
this was a highly atypical, subluminous event.  A few $JHK$ points
were obtained for SNe~1984A (Graham {\it et al.} 1988); 1989B (Kidger
{\it et al.} 1989; Wells {\it et al.}  1994) and 1991T (Menzies \&
Carter 1991; Harrison \& Stringfellow 1991). SN~1998bu in NGC~3368
(M96) was the first normal type~Ia since the 3~events described in
Elias {\it et al.}  (1981) for which reasonably good IR photometric
coverage was achieved (Jha {\it et al.}  1999, Meikle \& Hernandez
(1999), Hernandez {\it et al.} 2000).  In addition to the occurrence
of SN~1998bu, an important development since the work of Elias {\it et
al.} has been the establishment of more accurate distances to some of
the host galaxies.  Given these developments, it is timely to carry
out a re-examination of the extent to which SNe~Ia are homogeneous in
the IR.

\section{The sample}
I examine the IR ($JHK$) luminosities for eight SNe~Ia {\it viz.}  SNe
1972E, 1980N, 1981B, 1986G, 1989B, 1991T, 1998bu (Table~1).  These
were selected to have accurately known epochs of maximum blue light as
well as having reliable distance estimates and/or good light curve
coverage.  The photometric data are listed in the Appendix (Table~3).
Only measurements for which the quoted error was no greater than
$\pm$0.2~mag were included.  (Owing to its peculiarity and very sparse
IR coverage an exception to this rule was made for the peculiar SN~Ia
SN~1991T.)  Elias {\it et al.} (1985) also present IR photometry for
the type~Ia SNe~1971I, 71L, 83G, 83R, 83U, \& 84A, as do Graham {\it
et al.} (1988) for SN~1984A, but these were excluded on the basis of
the selection criteria.  In view of the rather unusual spectral energy
distribution of SNe~Ia in the $JHK$ bands, I make no attempt to
correct for differences in the detectors, filters or standards used
between different observatories.

\subsection{Distances}
The adopted distance moduli are given in Table~1, col.~4.  For
SNe~1972E, 1981B, 1989B \& 1998bu, HST-Cepheid distances to the host
galaxies are available.  For all except SN~1972E I use the
recalibrated Cepheid distance moduli given in Gibson {\it et al.}
(2000).  I show only their random errors in Table~1.  They also
estimate a systematic error of 0.16~mag, which would affect all their
distance moduli by the same amount.  The distance modulus to NGC~5253,
the host galaxy of SN~1972E, is still subject to quite substantial
uncertainty, with Saha {\it et al.} (1999) and Gibson {\it et al.}
(2000) disagreeing by 0.5~mag.  In view of this discrepancy, I have
adopted the mean of their values, {\it viz.} $27.86\pm0.25$ for
SN~1972E. \\

NGC~4527, the host galaxy of SN~1991T, lies close to NGC~4536 and
NGC~4496, both of which have HST Cepheid distances (Fisher {\it et
al.} (1999) and references therein).  I therefore adopt the mean
distance modulus of these two galaxies for NGC~4527.\\

SNe~1980N and 1981D both occurred in the same galaxy, NGC~1316
(Fornax~A).  Unfortunately there is not yet a Cepheid distance to this
galaxy, and the value of the distance is currently subject to some
disagreement.  Ciardullo, Jacoby \& Tonry (1993) use the planetary
nebula luminosity function (PNLF) distance indicator to obtain a
distance modulus of $31.19\pm0.07$ based on a distance modulus of
24.43 for M31.  A similar value for the Fornax Cluster is obtained by
Bureau, Mould \& Staveley-Smith (1996) who derive $31.10\pm0.45$ based
on an $I$-band Tully-Fisher distance relative to the Virgo Cluster,
and the HST-Cepheid distance to Virgo (Freedman {\it et al.} 1994).
Madore {\it et al.} (1999) have determined an HST-Cepheid distance to
the Fornax Cluster galaxy, NGC~1365, finding a modulus of
$31.35\pm0.07$.  They suggest that the distance to NGC~1316 is about
the same.  However, Saha {\it et al.} (1999) state that there are
reasons to suspect that NGC~1365 is in the {\it foreground} of the
Fornax Cluster.  Ferrarese {\it et al.} (2000) use their Cepheid-based
calibration of the $I$-band surface brightness fluctuation (SBF)
distance indicator to obtain a distance modulus to NGC~1316 of
$31.71\pm0.19$ and a Fornax Cluster distance modulus of
$31.59\pm0.04$.  They find that where $I$-band SBF and direct Cepheid
distance measurements to a galaxy are available good agreement is
obtained, while the PNLF distance indicator produces systematically
smaller distances.  (Ferrarese {\it et al.} also suggest that NGC~1365
is towards the front of the Fornax Cluster, similar to the conclusions
of Saha {\it et al.}).  Of course, one might use SN~1980N itself to
estimate its distance.  Hamuy {\it et al.} (1996b) obtain
$31.60\pm0.15$.  Equations (17) \& (18) in Phillips {\it et al.}
(1999) yield $31.70\pm0.08$, while Saha {\it et al.} (1999) quote a
value of $31.84\pm0.21$.  However, in a study such as this, where we
are attempting to investigate the homogeneity of SNe~Ia, it is
inadvisable to use distances determined from the phenomena we are
examining.  In view of the uncertainty in the distance to NGC~1316, I
have adopted the Ferrarese {\it et al.} (2000) Fornax Cluster
distance modulus of 31.59 for this galaxy.  However, given the typical
random errors given by Ferrarese {\it et al.} and the uncertainty of
the location of NGC~1316 within the Fornax Cluster, I assume a
somewhat larger error of $\pm$0.1~mag in the distance modulus of
NGC~1316. \\

There are no Cepheid distance measurements for NGC~5128
(Centaurus~A), the host galaxy of SN~1986G. Therefore I used the
weighted mean distance obtained from Cepheid-calibrated tip of the
red giant branch and $I$-band SBF methods (Ferrarese {\it et al.}
2000).  This gives a distance modulus of $28.01\pm0.12$ \\

\subsection{Extinction Correction} 
As indicated above, for most SNe~Ia there is disagreement as to how
much of the decline-rate-corrected colour distribution can be
attributed to dust extinction.  I have therefore investigated two
approaches to correcting the extinction.  In the first approach I have
followed the work of Phillips {\it et al.}  (1999) which provides
estimates of $E(B-V)$ for seven of the supernovae in my sample.  They
did not consider SN~1981D, but Mark Phillips kindly provided his
estimate for this supernova, using the same technique as in Phillips
{\it et al.} (1999).  He obtains a total $E(B-V)=0.18\pm0.09$.  The IR
data were then de-reddened assuming the extinction law of Cardelli,
Clayton \& Mathis (1989) with $R_V=3.1$.\\

In the second approach, no reddening corrections were made to the four
supernovae for which $(Bmax-Vmax)<0.2$.  For the four that exceeded
this value (SNe 1981D, 1986G, 1989B \& 1998bu) I made a correction for
extinction using the Cardelli, Clayton \& Mathis law.  However, I used
a lower value of $R_V$, {\it viz.} 2.6, which is the value found by
Phillips {\it et al.} (1999) from a fit to colour-magnitude plots for
events for which they derived non-zero $E(B-V)$.  The estimated total
extinctions are shown in Table~1, cols.~6--11.

\section{Light curves and colour evolution}
In this section, I examine the individual light curves and colours of
the eight supernovae.  For each supernova I converted the photometry
to intrinsic absolute magnitudes using the distance moduli and
extinctions given in Table~1.  It was found that the two approaches to
dealing with the extinction correction made little qualitative
difference to the conclusions.  Of course, a major reason for this IR
study was to try to reduce effects of extinction uncertainties.  As
would be expected, the luminosities derived by method~(1) were slightly
higher.  Further details are given in Section~4.  In the presentation
of the absolute light curves and colour evolution that follows,
de-reddening is by means of method~(2).

\subsection{Light Curves}
In Figures 1, 3 \& 5 I show the intrinsic absolute magnitude IR light
curves for the $J$, $H$ and $K$ bands covering the period up to
+110~days.  In order to display clearly the much greater density of
data at early times, I show the same light curves in Figures~2, 4 \&
6, but plotted out to just +60~days. Also plotted are the template IR
light curves of Elias {\it et al.} (1985).  These are based on $JHK$
light curves of SNe~1972E, 80N, 81B \& 81D plus a few points from
other SNe~Ia.  I adjusted the templates in epoch to give the best
match to SN~1980N.  This required Elias {\it et al.}'s fiducial
time,``$t_0=0$'', to be set to --6.25~days ({\it i.e.} 6.25~days {\it
before} t$_{Bmax}$).  I also truncated the templates at +2~days since
this is the earliest epoch for which Elias {\it et al.} (1981, 1985)
presented data.  The vertical positions of the templates are set at
the average absolute magnitude values for six of the events.  This is
discussed in detail in Section~4.  It can be seen that, with the
exception of SN~1986G, the templates provide a useful representation
of the IR light curves and that the dispersion in coeval magnitudes is
not great.  I now examine the IR light curves by era.

\subsubsection{$t<t_{Bmax}$}
There are four SNe for which data in this era are available {\it viz.}
SNe 1986G, 1989B, 1991T \& 1998bu.  The earliest-ever IR measurement
of a type~Ia supernova is the single $JHK$ observation of SN~1991T by
Menzies \& Carter (1991) at about --11~days.  This is discussed later. 

The observations of SN~1998bu show that SNe~Ia peak in the IR about
5~days {\it before} $t_{Bmax}$.  At $t=-5$~days we have data for
SNe~1986G, 1989B and 1998bu. SNe~1989B and 1998bu have a similar
$J$-magnitude of about --19.0 at --5~days. At $H$ \& $K$, SN1998bu
peaks at about --18.7 in $H$ and --18.8 in $K$, while SN~1989B is
brighter by about 0.25~mag.  Given the uncertainty in the distances of
SNe~1989B and 1998bu, this difference is not significant.  SN~1986G is
0.4--0.5~mag fainter than SN~1998bu.  Random errors in distance moduli
and extinction corrections are probably insufficient to account for
the apparent lower luminosity of SN~1986G.  I conclude that this
supernova was indeed significantly underluminous at this time.

\subsubsection{$t_{Bmax}<t<$+25~d}
Since the work of Elias {\it et al.} (1981) it has been known that
Type~Ia supernovae light curves exhibit {\it two} maxima in the $JHK$
bands.  A similar double peak is also seen in the $I$-band (Ford {\it
et al.} 1993).  In the $J$-band, a particularly pronounced mininum is
seen at $t=+15$~d ({\it cf.}  Figs.~1 \& 2).  In the $H$ and $K$-bands
(Figs.~3--6), a less pronounced minimum is seen, occurring somewhat
earlier, at about $t=+10$~d.  Interestingly, in spite of its peculiar
nature, the light curves of SN~1986G exhibit absolute magnitudes
similar to those of the other supernovae around +7~days.  Nothing can
be said about SNe~1972E or 1991T as no IR observations are available
during this time.

\subsubsection{+25~d$<t<$110~d}
For three of the four events for which we have post-+25~d light
curves, we can see that the decline rates and coeval absolute
magnitudes are similar.  The exception is SN~1986G which is fainter in
$JHK$ by typically one magnitude during much of this period.  It is
unlikely that such a large factor is the result of errors in the
estimates of distance or extinction.   

\subsection{Colours}
In Figs.~7 \& 8 I plot the $(J-H)$ and $(H-K)$ evolution,
respectively.  As above, I consider only the case in which the
extinction was corrected by method~(2). Also in Figs.~7 \& 8 I show
the colour evolution templates of Elias {\it et al.} (1985), again
truncated at day~+2. 

\subsubsection{$(J-H)$ colour}
The t=--11~d point of SN~1991T is at about $(J-H)=+0.6$.  However,
given that this is the only point at this epoch and that 91T had an
unusual optical spectrum at early times (Fisher {\it et al.} 1999),
such a red colour may not be typical.  Between --8~days and +2~days,
data exist for only SNe~1986G and 1998bu, plus SN~1989B at a single
epoch.  There is little evidence of colour change in this
period. SN~1998bu remains at $(J-H)\sim-0.2$ with SN~1986G somewhat
redder at $\sim-0.05$.  (It should be remembered that these colours
have already been corrected for extinction).  Between +2~d and +10~d
the behaviour of the six supernovae observed is quite uniform. At +4~d
the $(J-H)$ colour moves sharply to the red, reaching $(J-H)=+1.0$ by
day~+10.  Even SN~1986G is not strikingly different in its colour
evolution during this period.  Between +10 and +15~days, the general
trend is continued reddening, reaching $(J-H)=+1.3$ by day~+15. The
colour then moves to the blue reaching +0.65 by day~30.  After that
the colour reddens again, levelling out at about +1.6 by +65~d. The
post-+2~d evolution is essentially that described by Elias {\it et
al.} (1985).  The only clear exception to this behaviour is that
exhibited by SN~1986G.  It begins to move back to the blue at only
+10~days, and then to the red again at only +20~days.  SN~1986G seems
to mimic the $(J-H)$ behaviour of the other supernovae, but at a
higher rate of evolution, and never attaining quite the same amount of
redness.

\subsubsection{$(H-K)$ colour}
Again, the very early SN~1991T point shows an apparently unusually red
colour.  During the 8~days before $t_{Bmax}$, the $(H-K)$ evolution of
SN~1998bu shows a gradual reddening from --0.1 to +0.25.  After
$t_{Bmax}$, there is a gradual move to the blue, reaching
$(H-K)=-0.15$ at +40~days.  The behaviour of SN~1986G is again
different.  Up to about +50~days it is slightly bluer than typical.
From the very few post-+80~d observations, we note that SNe~1980N \&
1981B show a gradual drift to the red again.  In contrast, SN~1986G
shows a very striking move to the blue.  The unusually blue $(H-K)$
behaviour of SN~1986G was pointed out by Frogel {\it et al.} (1987).
The post-+2~d $(H-K)$ behaviour of the other supernovae is essentially
as described by Elias {\it et al.} (1981). 

The non-monotonic behaviour of the IR light curves and the development
of the IR colours have been considered by a number of authors.
H\"oflich, Khokhlov \& Wheeler (1995) have found that in some of their
models, the second maximum in the light curve is reproduced in some
regions of the IR. They attribute this to a time-dependent opacity
causing the effective emitting area to peak at a delayed time in the
IR compared with the optical region.  However, they have difficulty in
reproducing the strength of the effect as observed in all bands from
$I$ to $K$.  From analyses of IR spectra, Spyromilio, Pinto \& Eastman
(1994), and Wheeler {\it et al.} (1998) conclude that the appearance
of the red $(J-H)$ colour can be attributed to the relatively large
reduction of line-blanketing opacity in the $1.2-1.5~\mu$m region.
Wheeler {\it et al.} show that the depth of the $J$-band deficit can
provide a valuable temperature diagnostic for the silicon layers.

\section{Quantitative comparison of absolute magnitudes}
The presentation given above suggests that the IR behaviour of SNe~Ia
is generally quite homogeneous. I now endeavour to quantify this
behaviour.  A problem is that the coverage of IR light curves is still
much inferior to that achieved in the optical region.  In particular
we cannot, as optical studies do, compare directly the absolute
magnitudes at maximum light.  However, even in optical studies maximum
light is sometimes missed.  Nevertheless, the magnitude at maximum
light can still be estimated by fitting template light curves to the
data ({\it e.g.} Hamuy {\it et al.} 1995).  I have adopted a similar
approach for this study.  As pointed out in the previous section, the
IR light curve templates of Elias {\it et al} (1985) provide a fair
representation of the IR light curve shapes.  I have therefore used
these templates to compare the IR absolute magnitudes of the
supernovae. \\

I first compared the $H$-band template with the SN~1980N $H$-band
photometry and varied the position of the template in both axes
(absolute magnitude and time) to minimise $\chi_\nu^2$.  For the 14
points considered, a $\chi_\nu^2$ minimum of 1.7 was achieved by
setting $t_0=0$ at $-6.25\pm1.0$~days, where $t_0$ is the epoch as
defined by Elias {\it et al.} (1995).  This is consistent with the
value of --6~days adopted by Elias {\it et al.}  I fixed all three
templates such that $t_0=0$ is at --6.25~days.  I then adjusted the
vertical (absolute magnitude) position of the templates to provide a
minimum $\chi^2$-fit to the data for all supernovae and all 3 bands.
In spite of the fact that the templates were originally largely based
on SNe~1972E, 80N, 81B \& 81D, some of the best fits to these
particular SNe had values of $\chi_\nu^2$ as large as 10. However, it
was found that modest increases in the quoted errors on the data
points (up to $\pm$0.1~mag) would bring $\chi_\nu^2$ close to unity.
For the 10--12 points of SN~1998bu, the fits were less satisfactory,
yielding a $\chi_\nu^2$ exceeding 10.  Satisfactory fits were only
obtained by artificially increasing the errors to around
$\pm$0.15~mag.  The need to arbitrarily increase the individual errors
to achieve satisfactory fits probably indicates that the IR light
curves are not completely homogeneous in shape.  For SN~1989B good
fits were easily obtained for the 2 points which overlapped the
timespan of the templates.  For SN~1986G, in view of its clearly
different light curve shape in all three bands, no template fits were
performed.  Instead, the weighted mean values for the 4 points in the
+12 to +15 day period were used to estimate its $M_{14}(J)$,
$M_{14}(H)$ and $M_{14}(K)$ (defined below).  For SN~1991T only a
single epoch ($t=+55$~d) coincided with the template timespan and so
the templates were simply adjusted to coincide with the single point
at each waveband.  However, given the peculiar nature of SN~1991T, it
is possible that the templates are not representative of its true IR
light curves.  \\
 
In Table~2 I show the results of the template fitting procedure.  The
absolute magnitudes (cols. 2, 4 \& 6) are expressed in terms of
$M_{14}(J)$, $M_{14}(H)$ and $M_{14}(K)$, the absolute magnitudes at
$t=+13.75$~days. This epoch corresponds to the epoch of the
$H-H_{20}=0.0$ point on the original $H$-band template of Elias {\it
et al.} (1985). It lies close to the epoch of first minimum in the
$J$-band. The tabulated values are those obtained after increasing the
errors on the data points until $\chi_\nu^2$ was about unity.  (When
the fitting procedure was performed with the original quoted errors,
the absolute magnitudes returned never differed from the values shown
by more than a few percent.)  The numbers in parentheses give the
error in the smallest two decimal places of the $M_{14}$ values
returned by the $\chi^2$-fitting procedure.  The only exceptions are
SNe~1986G and 1991T where the errors in parentheses are based simply
on the published photometry errors (see above).  However, given the
peculiar nature of SN~1991T {\it and} the fact that the epoch of
observation is over a month beyond the fiducial $t=+13.75$~day epoch,
it is possible that the quoted errors on the $M_{14}$ values are
significantly underestimated for this supernova.  Columns 3, 5 \& 7
give the total random error in the absolute magnitude, comprising the
$\chi^2$-fitting or photometry error (shown in parentheses in cols.~2,
4 \& 6), random error in the distance ({\it cf.} Table~1), and the
extinction correction error. The error in the extinction correction
was taken to be half of the difference in extinction correction using
methods~(1) and (2). \\

Following the work of Phillips {\it et al.} (1999) and others, in
Figure~9 I compare the intrinsic absolute IR magnitudes at
$t=+13.75$~d, $M_{14}(J)$, $M_{14}(H)$, $M_{14}(K)$, with $\Delta
m_{15}(B)$, the decline in the $B$-band 15~days after maximum blue
light (Table~2, col.~8).  For six of the eight SNe (the exceptions are
SNe~1986G \& 1991T) the $M_{14}$ values are reasonably consistent with
zero difference between the events.  Single parameter fits to the
$M_{14}$ values for the six events yields $\chi_\nu^2$ values of 1.6,
1.8 and 1.4 in $J$, $H$ and $K$ respectively.  The actual values ({\it
i.e.} the weighted means) are $M_{14}(J)=-16.86\pm0.06$,
$M_{14}(H)=-18.22\pm0.05$ and $M_{14}(K)=-18.23\pm0.05$. These values
are indicated by the horizontal dotted lines in Fig.~9.  Henceforth,
these six supernovae will be referred to as the ``IR-normals''.  The
dispersion in all 3 bands is about 0.15~mag for the IR-normals.  As a
check, I fitted the templates to the light curves of all six
IR-normals simultaneously (about 60 points per band).  The same
$M_{14}$ values were obtained.  SN~1986G has similar $M_{14}(H)$ and
$M_{14}(K)$ to those of the IR-normals, but is is about 0.5~mag
brighter in $J$ than the IR-normal mean.  However, Figs.~1 to 6 show
that SN~1986G was clearly underluminous by a substantial amount before
$\sim$0~days and after $\sim$+25~days.  At later epochs
(post-+25~days), SN~1986G is fainter by as much as one magnitude in
the three bands.  SN~1991T appears to be overluminous but at a low
significance - $\sim 2\sigma$ in each band.  It should also be
recalled that SN~1991T has only a single set of $JHK$ observations (at
$t=+55~d$) which lies within the template temporal span. \\

Similar results are obtained when the entire analysis is repeated
using method~(1) for the extinction correction.  The mean $M_{14}$
values for the six IR normals are, respectively, 0.09, 0.04 and 0.02
mags brighter in $JHK$, with slightly smaller $\chi_\nu^2$ values {\it
viz.} 1.3, 1.5 and 1.2.

\section{Discussion and Conclusion}
For all three IR wavebands the template analysis indicates that six of
the eight supernovae considered have similar coeval magnitudes.
Indeed, to within the uncertainties, they are indistinguishable.  Of
the six, SNe~1972E, 1981B, 1989B, 1998bu and, probably, SN~1980N are
all spectroscopically ``Branch-normal'' (Branch, Fisher \& Nugent
1993).  SN~1981D cannot be classified in this way due to insufficient
spectroscopic coverage.  In contrast, the two exceptions, SNe~1986G
and 1991T, have long been recognised as spectroscopically peculiar.
Nevertheless, in spite of the peculiar behaviour of SN~1986G, there is
evidence that at around +7~days it had a similar IR luminosity to
that of the IR-normals.  \\

The IR-normals span $0.87<\Delta m_{15}(B)<1.31$. Over this range the
dispersion in absolute magnitude is about 0.15 in all three bands, and
this can be accounted for almost entirely by the uncertainties.  In
other words, at this level of uncertainty, the IR-normals show no
systematic variation in absolute $J$, $H$ or $K$ magnitude with
$\Delta m_{15}(B)$.  For $0.87<\Delta m_{15}(B)<1.31$ both Phillips
{\it et al.} (1999) and Saha {\it et al.} (1999) give peak absolute
magnitude ranges of about 0.35 in $B$, falling to about 0.2 in $I$. It
is possible that the absolute magnitude range continues to decline as
we move further into the IR, becoming too small to detect in $JHK$
with our limited data.  As a check, I examined the $B$ and $V$ band
absolute peak magnitudes specifically for the six IR-normals.  For SNe
1972E, 1981B, 1989B \& 1998bu I used the absolute magnitudes given in
Gibson {\it et al.}.  For SNe 1980N \& 1981D I used the $I$-band SBF
Fornax Cluster distance (Ferrarese {\it et al.} 2000) together with
the peak apparent magnitudes given in Hamuy {\it et al.} (1991).
Extinction corrections were made following both methods~(1) and (2) of
section 2.1.  In both the $B$ and $V$ bands I find no significant
trend of absolute $B$ or $V$ magnitude with $\Delta m_{15}(B)$.  The
dispersion is similar to that seen in the IR.  It is therefore likely
that the size of the sample considered here is simply too small and
the uncertainties in the absolute IR magnitudes too large to reveal a
$\Delta m_{15}(B)$ relation of even comparable strength to that found
in the optical region. \\

The IR magnitudes of SN~1991T at --11~days are $J=+11.73$, $H=+11.16$
and $K=+10.93$ (Menzies \& Carter 1991).  These translate to the
highest-ever IR-luminosities for a type~Ia event, especially in the
$H$ \& $K$ region ({\it cf.}  Figs. 1--6).  Although we lack coeval
points from other supernovae, it appears that SN~1991T was
exceptionally IR-luminous for a SN~Ia at this epoch.  Menzies \& Koen
(1991) report simultaneous $UBV$ photometry yielding $V=+12.4$,
$B-V=+0.11$ and $U-B=-0.71$.  Lira {\it et al.} (1998) report similar
$BVR$ magnitudes obtained just 0.3~d after the Menzies \& Carter IR
measurement, giving $V=+12.45$, $B-V=+0.09$ and $V-R=+0.06$.  Thus,
the IR magnitudes indicate a strong excess with respect to the optical
region {\it e.g.} $V-K\sim+1.5$.  The next IR observation of SN~1991T
was about 10~days later.  This took the form of a $JHK$ spectrum
(Meikle {\it et al.} 1996), and it shows no evidence of excessive IR
luminosity.  The --11~day photometry of SN~1991T was the earliest ever
observation of a SN~Ia in the IR.  The next earliest are the IR
spectra of SN~1994D (Meikle {\it et al.} 1996) and photometry of
SN~1998bu, both at about --8.5~d. They show no signs of unusually high
luminosity.  It may be that the exceptional early-time IR luminosity
and redness of SN~1991T are further manifestations of its intrinsic
peculiarity.  However, for the present, this result is something of a
puzzle. \\

The analysis presented here indicates that most SNe~Ia are {\it
standard} IR candles at a level of about 0.15 mag.  As already
mentioned, Elias {\it et al.} (1995) suggested that SNe~Ia might have
a dispersion in IR absolute magnitude as low as $\pm$0.1~mag.  The
present work is consistent with this, although it also shows there can
be exceptional events which are much brighter or fainter than the
norm.  Moreover, although I find no indication of differences in
coeval magnitudes for the six IR-normals, there is tentative evidence
that the IR light curve {\it shapes} are not exactly identical.\\

It should be kept in mind that this study makes use of only eight
supernovae, and that their light curves are generally much less
well-sampled than in the optical region.  The {\it total} number of
photometry points for all six IR-normals is only about 60 in each band
within the span of the Elias {\it et al} templates.  It is clearly
desirable to pursue the use of SNe~Ia IR light curves as distance
indicators.  As a first step, we need to build up a good set of IR
light curves for nearby SNe~Ia in order to establish the extent to
which they are homogeneous.  In addition, this will extend the
wavelength range over which colour information is available and so
should help to distinguish unambiguously between intrinsic colours and
dust reddening.  Compared with the optical region, the lower
sensitivity to extinction uncertainties in the IR should mean that
observations in this wavelength region will provide more reliable
estimates of cosmological distances.  At the lowest redshifts
dominated by the Hubble expansion, ($z=0.01-0.1$), the peak IR
magnitudes of SNe~Ia would lie in the range +14 to +20.  These
magnitudes are well within the range of a 4~m class telescope such as
UKIRT ({\it cf.}
http://www.jach.hawaii.edu/JACpublic/UKIRT/\\instruments/ufti/sensitivities.html).
Indeed, the nearer supernovae could be successfully monitored by only
a 2~m class telescope such as the Liverpool Telescope ({\it cf.}
http://telescope.livjm.ac.uk/inst/index.html).  At higher redshifts,
say 0.5, rest frame $J$-band and $H$-band emission are shifted into
roughly the $H$- and $K$-bands respectively. (Rest frame $K$-band
emission would be lost in the thermal IR region.)  In a flat, $\Lambda
= 0$ universe an intrinsic peak absolute $J$-magnitude of --19.0
yields an apparent magnitude of about +22.3.  This estimate includes
only an approximate K-correction (Poggianti 1997).  As more SNe~Ia IR
spectra become available, it should be possible to make more accurate
estimates of the K-corrections.  A $J\sim+23$ SN~Ia at $z=0.44$ has
been successfully detected by J.~Spyromilio in a 1-hour integration
with the 3.5~m ESO NTT (Riess et al., in preparation). Clearly, even
deeper coverage would be possible with an 8~m class telescope such as
Gemini ({\it cf.}
http://www.ast.cam.ac.uk/sciops/instruments/niri/NIRI\\Index.html).

\section*{Acknowledgements}
I am very grateful to Mark Phillips for his assistance with the
parameters for SN~1981D and for providing his estimate of the
$M_{Bmax}$ against $(Bmax-Vmax)$ slope for SNe~Ia with
$(Bmax-Vmax)<0.2$.  I also thank Alexandra Fassia, Miguel Hernandez,
Bruno Leibundgut, Mark Phillips, and Michael Rowan-Robinson for
commenting on an ealier version of this paper.

\begin{table*}
\centering
\caption[]{Supernova Sample}
\begin{minipage}{\linewidth}
\renewcommand{\thefootnote}{\thempfootnote}
\begin{tabular}{rccccccccccc} \hline
Supernova & 
t$_{Bmax}$\footnote{Epoch sources - SN~1972E: Leibundgut {\it et al.} (1991), SNe~1980N \& 1981D: Hamuy {\it et al.} (1991),
SN~1981B: Shaefer (1995),\\ SN~1986G: Phillips {\it et al.} (1987), SN~1989B: Wells {\it et al.} (1994), SN~1991T: Lira {\it
et al.} (1998), SN~1998bu: Hernandez {\it et al.} (2000)}
(d)\footnote{Julian Date -- 2400000.} & 
Host Gal. & 
$\mu$\footnote{Distance Modulus: the error in parentheses is the random error only - see text.} &
$E(B-V)$\footnote{$E(B-V)$ for all SNe except SN~1981D is from Phillips {\it et al.} (1999) and includes Galactic reddening.  $E(B-V)$ for SN~1981D is from M.~Phillips (private communication).} &  
\multicolumn{2}{c}{$A_J$\footnote{The IR extinctions were determined using two methods: (1) \& (2) - see text.}} &  
\multicolumn{2}{c}{$A_H$} & 
\multicolumn{2}{c}{$A_K$} & 
Source\footnote{IR photometry obtained from:\\ 
\indent 1. Elias {\it et al.} (1985),
2. Elias {\it et al.} (1981),
3. Frogel {\it et al.} (1987),
4. Kidger {\it et al.} (1989),
5. Wells {\it et al.} (1994),\\
\indent 6. Menzies \& Carter (1991), 
7. Harrison \& Stringfellow (1991),
8. Mayya, Puerari \& Kuhn (1998),
9. Jha {\it et al.} (1999),\\
\indent 10. Meikle \& Hernandez (1999); Hernandez {\it et al.} (2000).} \\ \hline
&&&&&(1)&(2)&(1)&(2)&(1)&(2)&\\ \hline
1972E & 41449.0(1.0)\footnote{Numbers in parentheses give random error in the smallest two decimal 
places.}
& NGC 5253  & 27.86(25) & 0.07(04)    & 0.06 & 0.0 & 0.04 & 0.0 & 0.02 & 0.0 & 1 \\
1980N & 44585.8(0.5) & NGC 1316  & 31.59(10) & 0.07(05)  & 0.06 & 0.0 & 0.04 & 0.0 & 0.02 & 0.0 & 2 \\
1981B & 44672.0(0.2) & NGC 4536  & 30.95(07) & 0.13(03)  & 0.11 & 0.0 & 0.07 & 0.0 & 0.04 & 0.0 & 1,2 \\
1981D & 44679.9(0.5) & NGC 1316  & 31.59(10) & 0.18(06)  & 0.16 & 0.13 & 0.10 & 0.08 & 0.06 & 0.05 & 2 \\
1986G & 46561.5(1.0) & NGC 5128  & 28.01(12) & 0.61(05) & 0.54 & 0.45 & 0.34 & 0.28 & 0.22 & 0.18 & 3 \\
1989B & 47565.3(1.0) & NGC 3627  & 30.06(17) & 0.37(04) & 0.32 & 0.27 & 0.20 & 0.17 & 0.13 & 0.11 & 4,5 \\
1991T & 48375.7(0.5) & NGC 4536  & 31.07(13) & 0.16(05) & 0.14 & 0.0 & 0.09 & 0.0 & 0.06 & 0.0 & 6,7 \\
1998bu & 50953.3(0.5) & NGC 3368 & 30.20(10) & 0.35(03) & 0.31 & 0.26 & 0.19 & 0.16 & 0.12 & 0.10 & 8,9,10 \\
\hline
\end{tabular} \\
\end{minipage}
\end{table*}

\vspace{2cm}

\begin{table*}
\centering
\caption[]{Absolute magnitudes at $t=+13.75$~days.}
\begin{minipage}{\linewidth}
\renewcommand{\thefootnote}{\thempfootnote}
\begin{tabular}{rccccccc} \hline
SN & $M_{14}(J)$\footnote{Numbers in parentheses give the error in the smallest
two decimal places of $M_{14}(J)$, $M_{14}(H)$ and $M_{14}(K)$ returned by the $\chi^2$-fitting procedure.} 
& $\Delta M_{14}(J)$\footnote{Total random error in the absolute magnitude.  It 
includes $\chi^2$-fitting error, random error in the distance estimates ({\it cf.} Table~1), and extinction correction error (see text).}
& $M_{14}(H)$ & $\Delta M_{14}(H)$ & $M_{14}(K)$ & $\Delta M_{14}(K)$ &
$\Delta m_{15}(B)$\footnote{Values of $\Delta m_{15}(B)$ taken from Phillips 
{\it et al.} (1999), apart from SN~1981D. For this supernova the value was 
provided by M. Phillips (private communication). Numbers in parentheses give the error in the smallest two decimal places.} 
\\ \hline
1972E  & --17.26(16) & 0.30  & --18.52(04) & 0.25  &  --18.64(06) & 0.25 & 0.87(10) \\
1980N  & --17.02(03) & 0.11  & --18.29(01) & 0.10  &  --18.33(01) & 0.10 & 1.28(04) \\
1981B  & --16.70(04) & 0.10  & --18.03(03) & 0.09  &  --18.09(02) & 0.08 & 1.10(07) \\
1981D  & --16.94(05) & 0.11  & --18.32(03) & 0.10  &  --18.24(02) & 0.10 & 1.27(09) \\
1986G  & --17.39(04) & 0.14  & --18.27(02) & 0.13  &  --18.18(02) & 0.13 & 1.73(07) \\
1989B  & --16.71(04) & 0.18  & --18.03(06) & 0.18  &  --18.24(13) & 0.21 & 1.31(07) \\
1991T\footnote{Given the peculiar nature of SN~1991T {\it and} the
fact that the epoch of observation is over a month beyond the fiducial
$t=+13.75$~day epoch, it is possible that the quoted errors on
$M_{14}(J)$, $M_{14}(H)$ and $M_{14}(K)$ are significantly
underestimated for this supernova.}
& --17.47(30) & 0.33  & --18.63(14) & 0.20  &  --18.78(25) & 0.28 & 0.94(05) \\
1998bu & --16.81(07) & 0.13  & --18.29(03) & 0.11  &  --18.27(04) & 0.11 & 1.01(05) \\
\hline
\end{tabular} \\
\end{minipage}
\end{table*}

\begin{table*}
\centering
\caption[]{Appendix: Infrared photometry of type Ia supernovae}
\begin{minipage}{\linewidth}
\renewcommand{\thefootnote}{\thempfootnote}
\begin{tabular}{llllll} \hline \hline
Supernova & $t_{Bmax}$(d)\footnote{JD -- 2400000} & 
$J$\footnote{Numbers in parentheses give the error in the smallest one or two decimal places.}  & 
$H$ & $K$ & Source  \\ \hline
SN 1972E & 41459.7   &      --      &    --      &   9.44(20) & Elias {\it et al.} (1985) \\
& 41461.8   &      --      &    --      &   9.29(16) & \\
& 41463.6   &      --      &   9.32(8)  &   9.35(8)  & \\
& 41472.6   &    10.37(12) &   9.07(8)  &   9.18(8)  & \\
& 41482.8   &      --      &    --      &   9.00(16) & \\
& 41487.7   &    10.76(10) &   9.76(8)  &   9.95(8)  & \\
& 41488.7   &      --      &   9.83(10) &   9.96(9)  & \\
& 41493.7   &      --      &  10.19(10) &  10.35(10) & \\
& 41496.7   &    11.16(13) &  10.39(10) &  10.24(11) & \\
& 41502.7   &      --      &  10.61(17) &    --      & \\
& 41516.7   &      --      &  11.16(16) &  11.40(17) & \\
& 41517.7   &      --      &  11.17(11) &  11.29(8)  & \\
& 41518.7   &    12.58(16) &    --      &  11.03(13) & \\ 
\hline
SN 1980N & 44590.7   &    13.32(5)  &  13.43(5)  &  13.23(8)  & Elias {\it et al.} (1981, 1985) \\
& 44591.5   &    13.50(7)  &  13.46(8)  &  13.32(9)  & \\
& 44592.7   &    13.61(4)  &  13.38(2)  &  13.32(5)  & \\ 
& 44596.6   &    14.40(4)  &  13.44(3)  &  13.38(4)  & \\
& 44618.6   &    14.15(3)  &  13.47(2)  &  13.59(2)  & \\
& 44621.6   &    14.47(3)  &  13.66(2)  &  13.82(2)  & \\
& 44624.6   &    14.76(4)  &  13.84(2)  &  14.01(3)  & \\
& 44631.7   &    15.42(5)  &  14.18(3)  &  14.34(4)  & \\
& 44653.5   &    16.74(8)  &  15.11(4)  &  15.20(11) & \\
& 44657.6   &      --      &  15.15(8)  &    --      & \\
& 44667.5   &      --      &  15.74(11) &    --      & \\
& 44676.5   &    17.65(16) &  16.00(8)  &    --      & \\
& 44677.5   &    17.69(20) &  16.24(7)  &  16.21(11) & \\
& 44685.5   &    17.54(19) &  16.50(9)  &    --      & \\ 
\hline
SN 1981B & 44674.67  & 12.70(5) &  13.00(5) & 12.60(5) & Elias {\it et al.} (1981, 1985) \\  
& 44679.0    &   13.19(7)  &  12.95(6)  &  12.81(5)  & \\      
& 44680.73  &    13.75(5)  &  13.27(5)  &  12.99(5)  & \\
& 44685.7   &    14.18(3)  &  12.98(2)  &  12.89(2)  & \\
& 44687.7   &    14.20(3)  &  12.90(2)  &  12.90(2)  & \\   
& 44689.0   &    14.29(2)  &  12.93(3)  &  12.87(4)  & \\
& 44689.8   &    14.11(3)  &  12.84(2)  &  12.67(2)  & \\
& 44690.8   &    14.17(2)  &  12.82(2)  &  12.81(3)  & \\
& 44691.7   &    13.85(3)  &  12.69(2)  &  12.77(2)  & \\
& 44692.8   &    13.99(3)  &  12.73(5)  &  12.81(4)  & \\
& 44695.8   &    13.66(3)  &  12.59(3)  &  12.68(5)  & \\
& 44708.7   &    14.01(11) &  13.19(8)  &    --      & \\
& 44737.7   &    16.09(15) &  14.53(7)  &  14.67(7)  & \\
& 44738.5   &    16.19(6)  &  14.53(3)  &  14.67(3)  & \\
& 44748.5   &    16.65(5)  &  15.02(3)  &  15.03(5)  & \\
& 44768.5   &    17.52(11) &  15.80(4)  &  15.71(5)  & \\
& 44782.5   &    17.87(13) &  16.22(8)  &  15.99(13) & \\
\hline
SN 1981D & 44683.5  &    13.38(3) &   13.41(3) &  13.36(4) & Elias {\it et al.} (1981, 1985) \\
& 44685.5   &    13.74(2)  &  13.45(2)  &  13.43(3)  & \\
& 44686.5   &    14.05(2)  &  13.56(2)  &  13.55(3)  & \\
& 44687.5   &    14.27(3)  &  13.56(2)  &  13.57(3)  & \\
& 44688.5   &    14.44(3)  &  13.59(3)  &  13.58(4)  & \\
& 44693.5   &    14.58(4)  &  13.28(3)  &  13.28(7)  & \\
& 44695.5   &    14.49(6)  &  13.22(3)  &  13.25(6)  & \\
\hline
\end{tabular} \\
\end{minipage}
\end{table*}
\begin{table*}
\centering
\begin{minipage}{\linewidth}
\renewcommand{\thefootnote}{\thempfootnote}
\begin{tabular}{llllll} 
\multicolumn{6}{l}{{\bf Table 3} {\it (continued)}. Appendix. Infrared photometry of type Ia supernovae} \\
\hline \hline
Supernova & $t_{Bmax}$(d)\footnote{JD -- 2400000} & 
$J$\footnote{Numbers in parentheses give the error in the smallest one or two decimal places.} & 
$H$ & $K$ & Source  \\ \hline

SN 1986G & 46556.4  & 10.10(3) & 9.96(3)  & 9.90(3)  & Frogel {\it et al.} (1987) \\
& 46557.4   &    10.06(3)  &   9.97(3)  &   9.87(3)  & \\
& 46562.2   &    10.12(3)  &   9.99(3)  &   9.87(3)  & \\
& 46563.4   &    10.28(3)  &  10.10(3)  &  10.03(3)  & \\
& 46565.8   &      --      &  10.10(3)  &  10.02(3)  & \\
& 46567.3   &    10.73(3)  &   9.99(3)  &  10.03(3)  & \\
& 46568.3   &    10.86(3)  &  10.02(3)  &  10.05(3)  & \\
& 46568.4   &    10.81(3)  &  10.03(3)  &  10.08(3)  & \\
& 46569.3   &    10.88(3)  &   9.96(3)  &  10.01(3)  & \\
& 46569.3   &    10.94(3)  &   9.95(3)  &   9.99(3)  & \\
& 46570.7   &    11.04(3)  &   9.97(3)  &   9.98(3)  & \\
& 46571.4   &    11.01(3)  &  10.00(3)  &   9.92(3)  & \\
& 46573.4   &    11.01(3)  &  10.00(3)  &   9.99(3)  & \\
& 46575.5   &    11.11(3)  &  10.03(3)  &  10.08(3)  & \\
& 46575.5   &    11.15(5)  &  10.00(5)  &   9.99(5)  & \\
& 46576.5   &    11.00(3)  &  10.03(3)  &   9.97(3)  & \\
& 46580.5   &    11.04(5)  &  10.17(5)  &  10.15(5)  & \\
& 46581.3   &    10.99(3)  &  10.21(3)  &  10.20(3)  & \\
& 46581.4   &    10.95(3)  &  10.23(3)  &  10.27(3)  & \\
& 46581.5   &    11.09(5)  &  10.21(5)  &  10.21(5)  & \\
& 46582.5   &    11.16(5)  &  10.29(5)  &  10.29(5)  & \\
& 46583.5   &    11.23(3)  &  10.39(3)  &  10.42(3)  & \\
& 46584.5   &    11.35(3)  &  10.49(3)  &  10.53(3)  & \\
& 46585.5   &    11.50(3)  &  10.60(3)  &  10.65(3)  & \\
& 46587.5   &    11.92(8)  &  10.85(3)  &  10.90(3)  & \\
& 46596.4   &    12.42(3)  &  11.27(3)  &  11.39(3)  & \\
& 46607.3   &    13.15(3)  &  11.75(3)  &  11.97(3)  & \\
& 46638.2   &    14.50(10) &  13.10(3)  &  13.52(3)  & \\
& 46662.5   &    14.62(12) &  13.56(7)  &  14.10(7)  & \\
\hline

SN 1989B & 47560.5 & 11.28(3) & 11.21(3) & 11.04(3)  & Kidger {\it et al.} (1989) \\
& 47571.8   &    12.80(7)  &  12.40(7)  &  12.20(11) & Wells {\it et al.} (1994) \\
& 47584.9   &    13.22(5)  &  11.91(5)  &  11.58(10) & `` \\
\hline
SN 1991T & 48364.4 & 11.73(3) & 11.16(3) & 10.93(3) & Menzies \& Carter (1991) \\
& 48429.9   &   14.97(30)     &  13.72(14)  &  13.72(25)  & Harrison \& Stringfellow (1991) \\
\hline
SN 1998bu & 50944.8 & 11.77(5) & 11.73(5) & 11.81(5) & Mayya, Puerari \& Kuhn (1998) \\
& 50945.6   &    11.76(6)  &  11.88(6)  &  11.81(5)  & Jha {\it et al.} (1999) \\
& 50945.7   &    11.67(5)  &  11.66(5)  &  11.67(5)  & Hernandez {\it et al.} (2000) \\  
& 50947.7   &    11.49(5)  &  11.58(5)  &  11.42(5)  & \\
& 50948.6   &    11.59(6)  &  11.77(6)  &  11.59(5)  & \\
& 50949.4   &    11.55(3)  &  11.59(3)  &  11.42(3)  & \\
& 50950.4   &    11.68(3)  &  11.86(3)  &  11.44(3)  & \\
& 50951.9   &    11.66(4)  &  11.84(4)  &  11.60(3)  & \\
& 50952.7   &    11.71(4)  &  11.88(4)  &  11.63(3)  & \\
& 50953.4   &    11.89(5)  &  11.95(5)  &  11.60(5)  & \\ 
& 50955.4   &    11.87(5)  &  11.83(5)  &  11.66(5)  & \\ 
& 50957.4   &    12.06(5)  &  11.88(5)  &  11.61(5)  & \\ 
& 50958.4   &    12.05(5)  &  11.96(5)  &  11.75(5)  & \\ 
& 50959.8   &    12.42(3)  &  11.97(3)  &  11.88(3)  & \\ 
& 50970.3   &      --      &  11.75(5)  &  11.84(4)  & \\
& 50970.7   &    13.32(6)  &  11.94(5)  &  11.95(5)  & \\
& 50974.9   &    13.23(6)  &  11.68(6)  &  11.89(5)  & \\
& 50976.0   &    13.12(4)  &  11.79(3)  &  11.77(3)  & \\
& 50976.8   &    13.08(1)  &  11.74(2)  &  11.92(2)  & \\
& 50976.9   &    13.08(6)  &  11.65(6)  &  11.77(5)  & \\
& 50978.6   &    12.81(6)  &  11.73(6)  &  11.74(5)  & \\
& 50978.9   &      --      &  11.67(10) &  11.77(10) & \\
& 50984.8   &    12.68(5)  &  12.00(3)  &  12.05(4)  & \\
\hline
\end{tabular} \\
\end{minipage}
\end{table*}

\begin{figure*}
\vspace{13.0cm}
\includegraphics{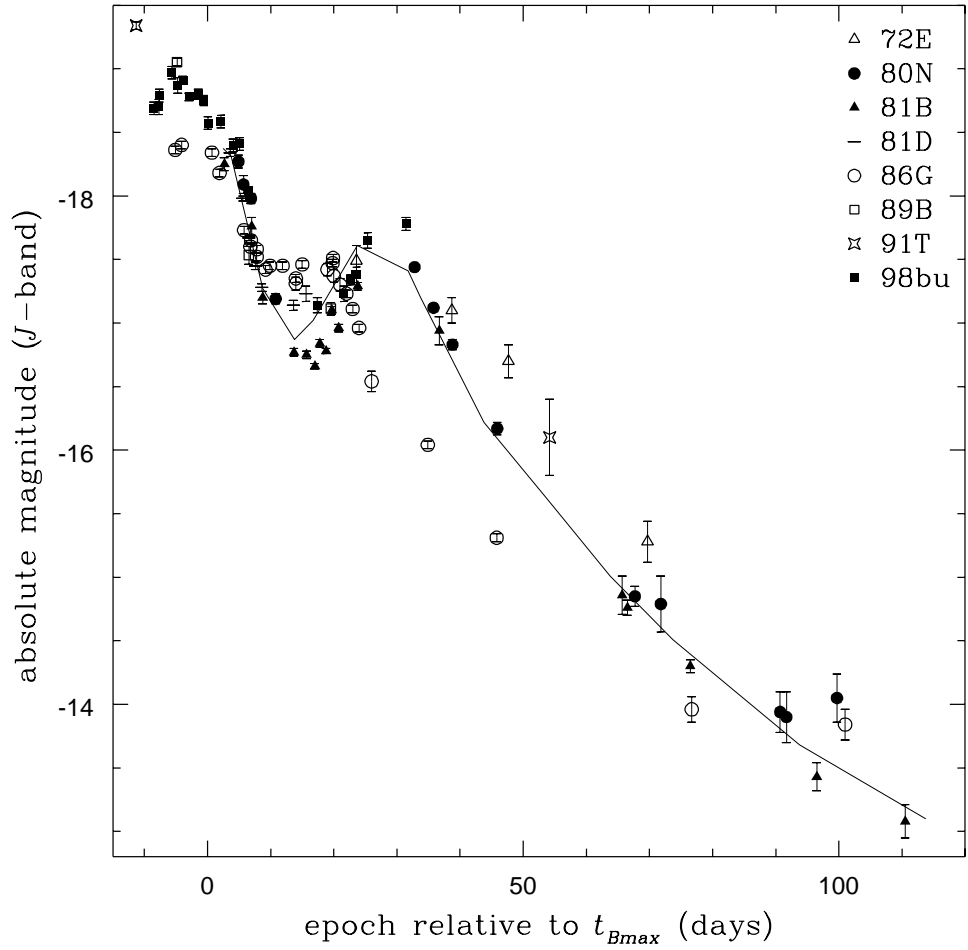}
\caption[]{$J$-band light curves of type~Ia supernovae.  The absolute
magnitudes were derived as explained in the text, using method~(2) to
correct for the extinction. The error bars give the photometry errors
only. The data set for each supernova is also subject to uncertainty
in distance and extinction correction.  The continuous line is the
template light curve of Elias {\it et al.} (1985) with their
$t_0=0$ set at --6.25~days, and with $M_{14}(J)=-16.86$ (see text).}
\end{figure*}

\begin{figure*}
\vspace{13.0cm}
\includegraphics{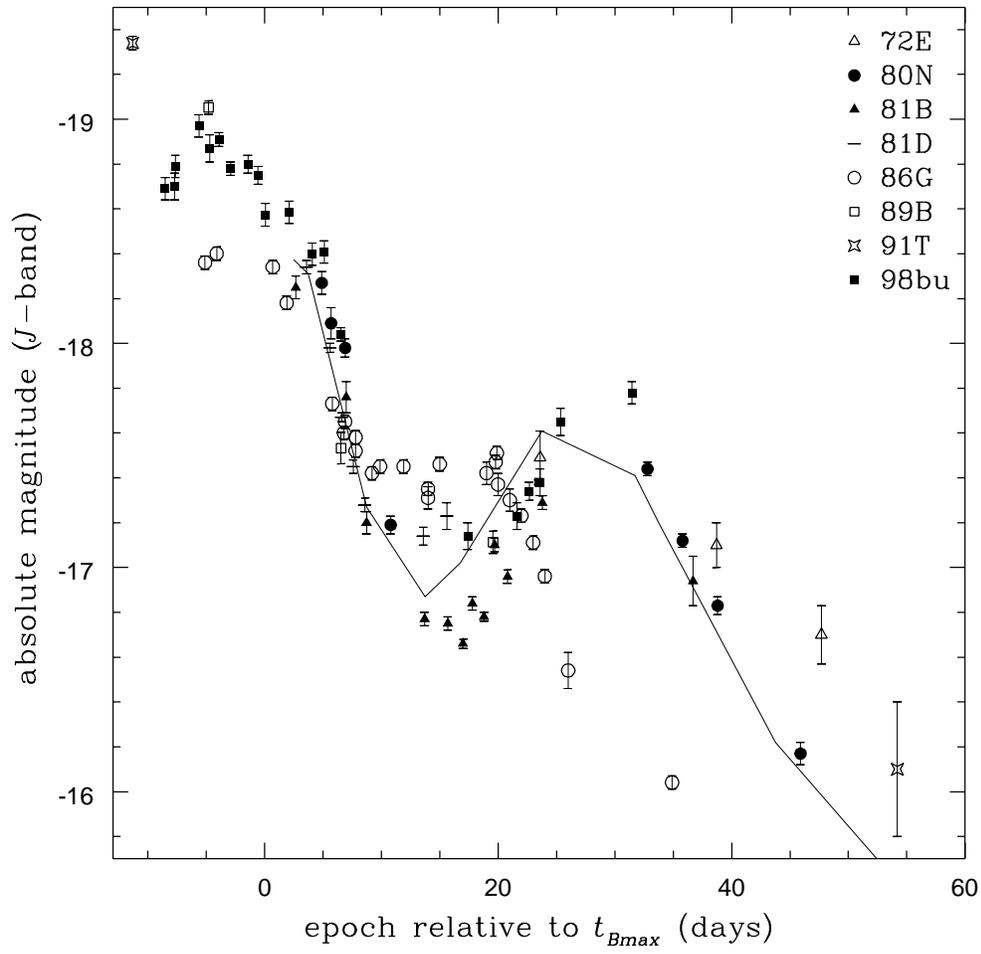}
\caption[]{Detail of Figure~1.}
\end{figure*}

\begin{figure*}
\vspace{13.0cm}
\includegraphics{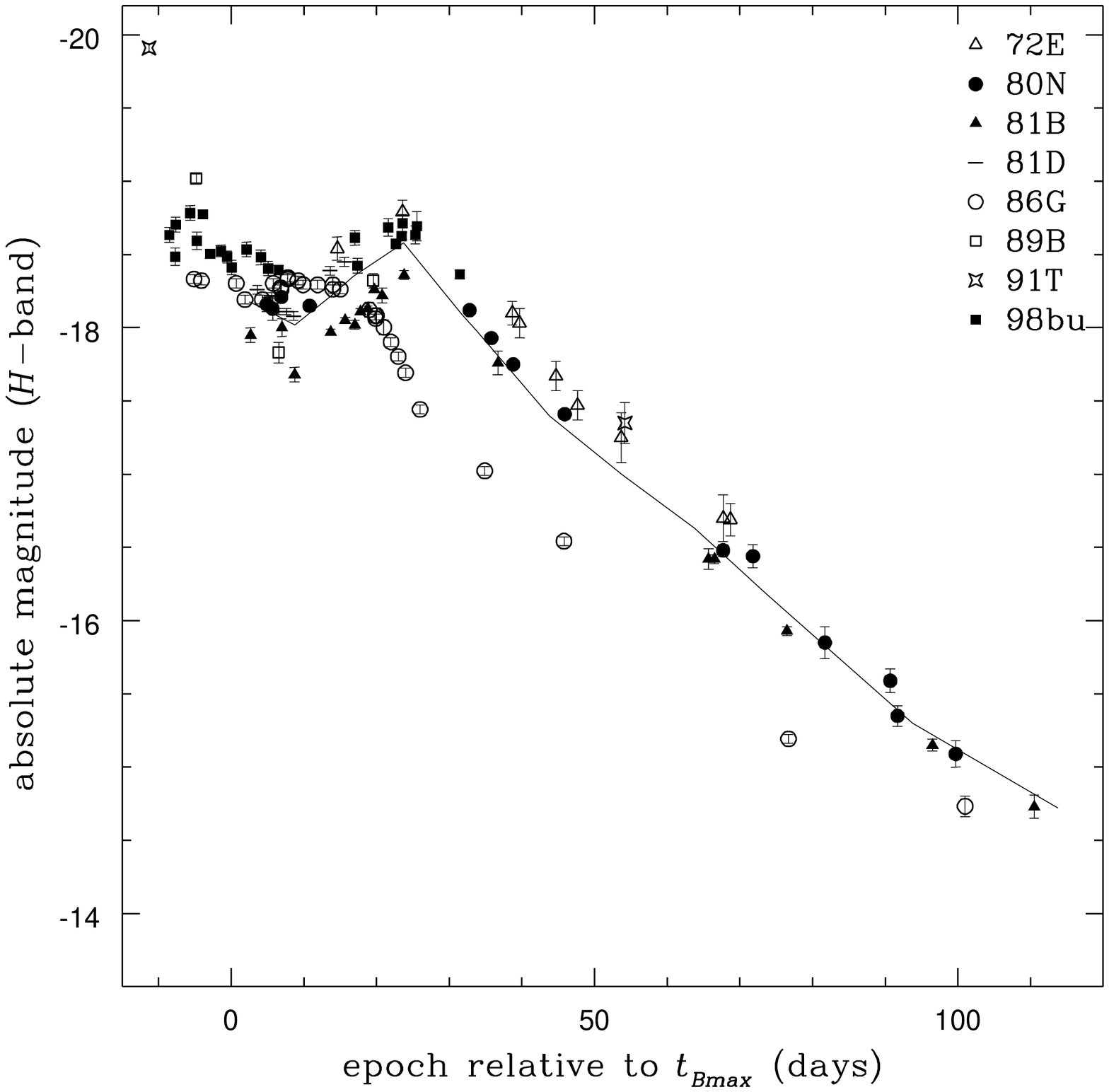}
\caption[]{$H$-band light curves of type~Ia supernovae.  The template
light curve has $M_{14}(H)=-18.22$.  Other details as in Fig.~1
caption.}
\end{figure*}

\begin{figure*}
\vspace{13.0cm}
\includegraphics{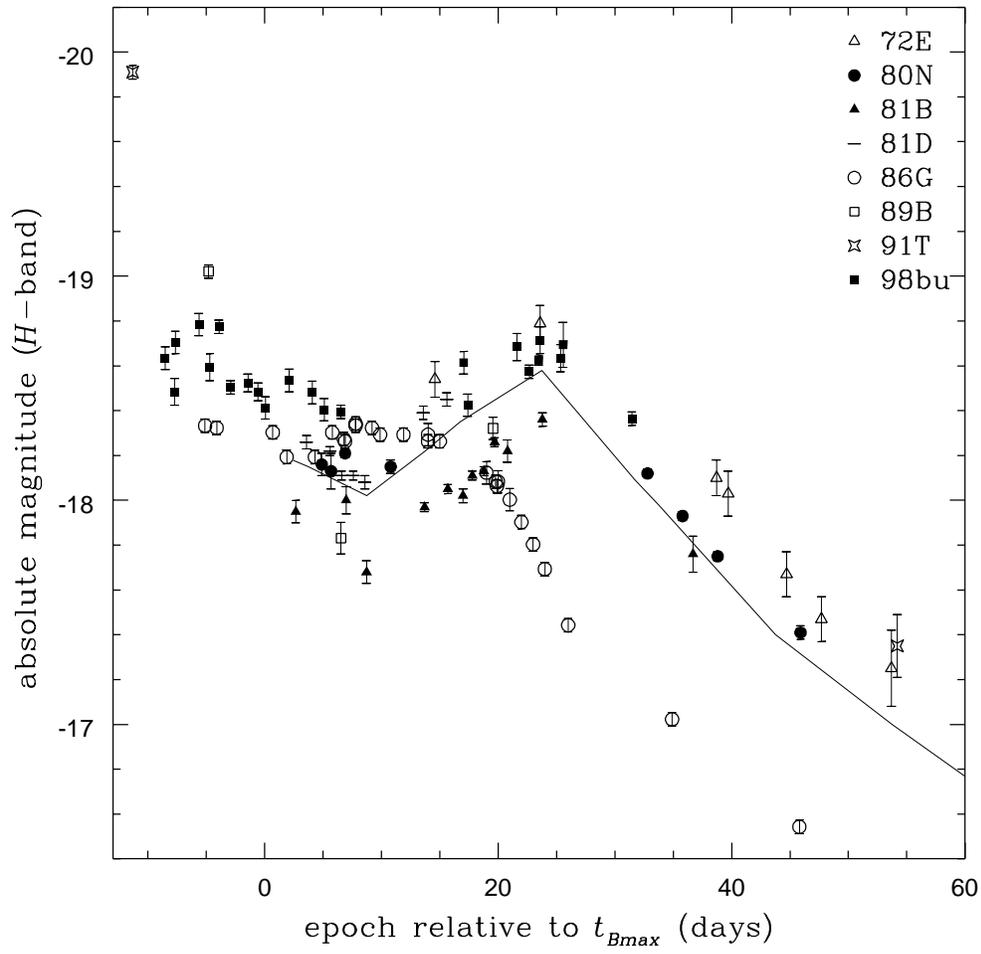}
\caption[]{Detail of Figure~3.}
\end{figure*}

\begin{figure*}
\vspace{13.0cm}
\includegraphics{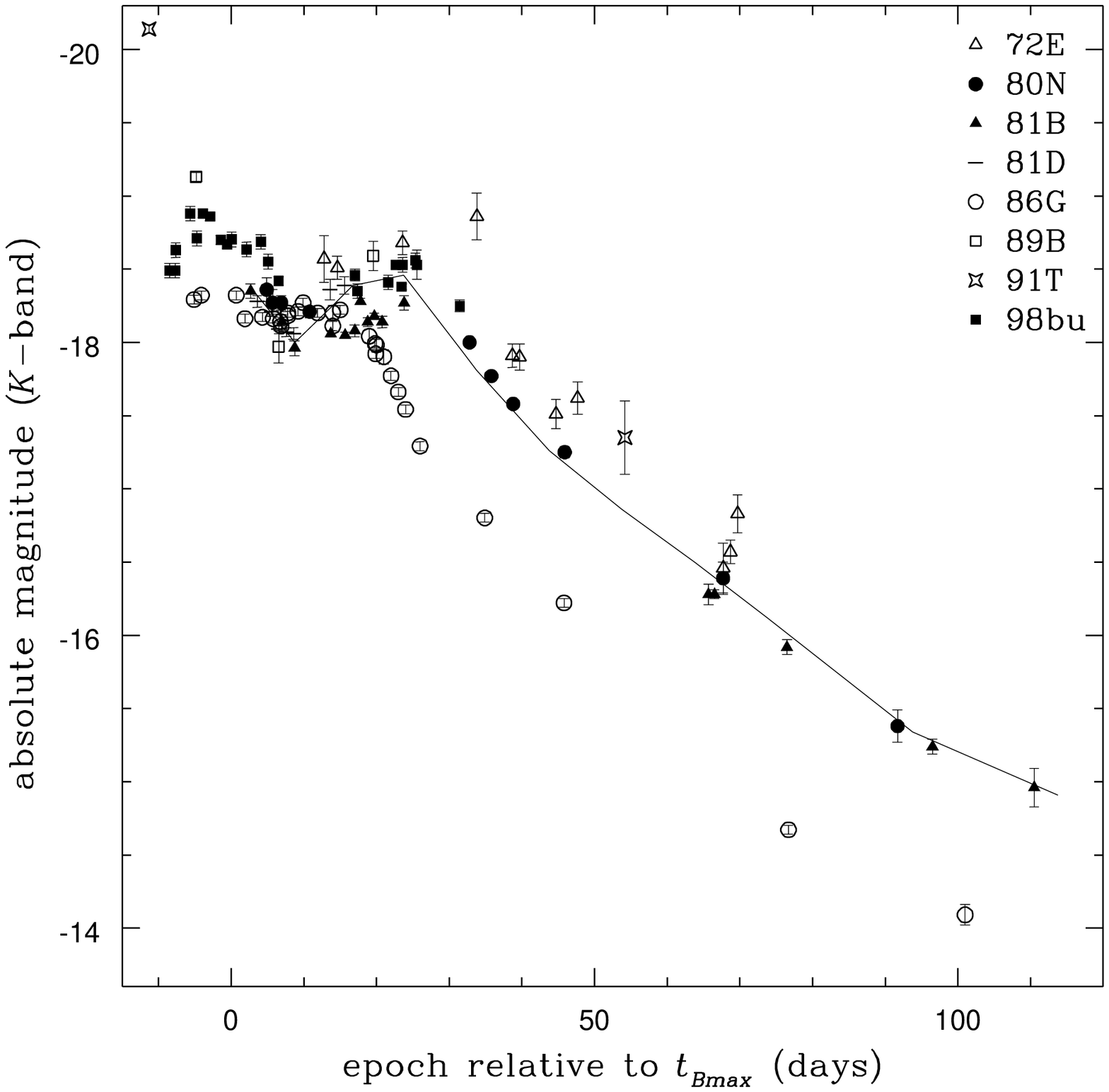}
\caption[]{$K$-band light curves of type~Ia supernovae.  The template
light curve has $M_{14}(K)=-18.23$.  Other details as in Fig.~1
caption.}
\end{figure*}

\begin{figure*}
\vspace{13.0cm}
\includegraphics{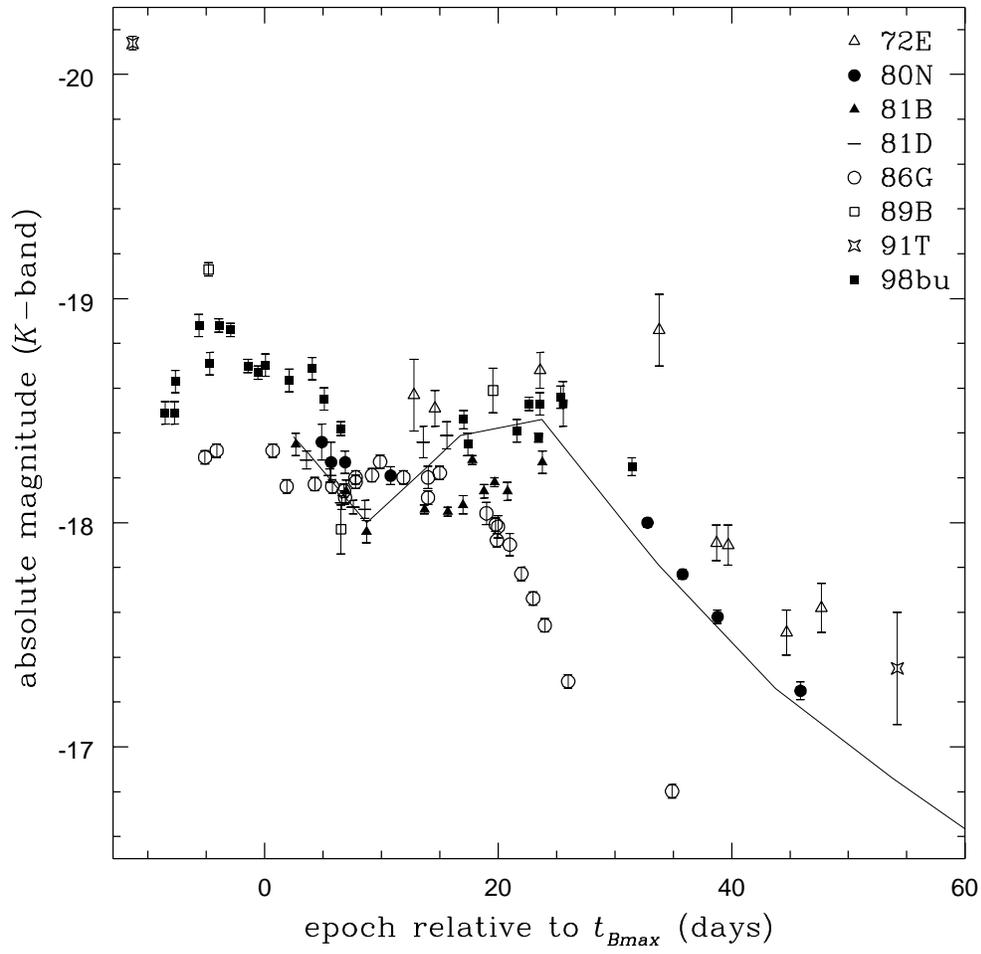}
\caption[]{Detail of Figure~5.}
\end{figure*}

\begin{figure*}
\vspace{13.0cm}
\includegraphics{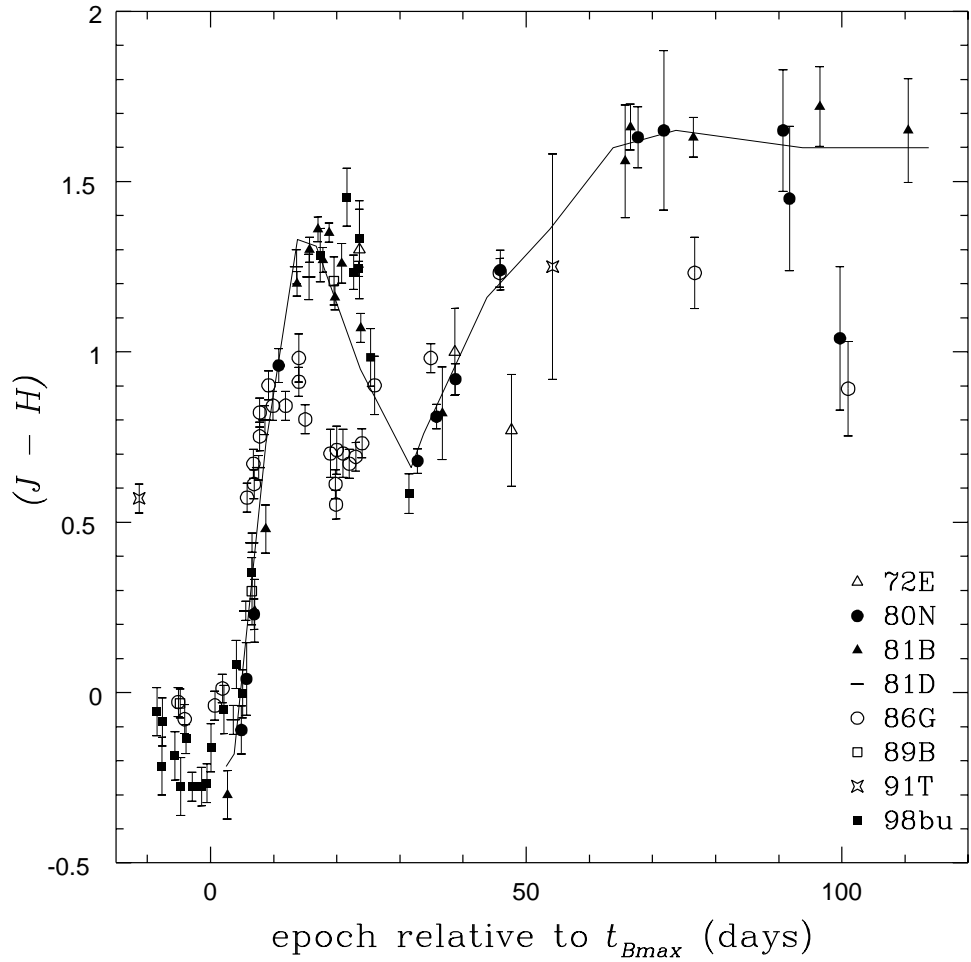}
\caption[]{Evolution of the $(J-H)$ colour of type~Ia supernovae.  The
colours have been de-reddened by method~(2) as described in the text.
The error bars represent random error only. The data set for each
supernova is also subject to uncertainty in extinction correction.
The continuous line is the template light curve of Elias {\it et al.} (1985)
with their $t_0=0$ set at --6.25~days.}
\end{figure*}

\begin{figure*}
\vspace{13.0cm}
\includegraphics{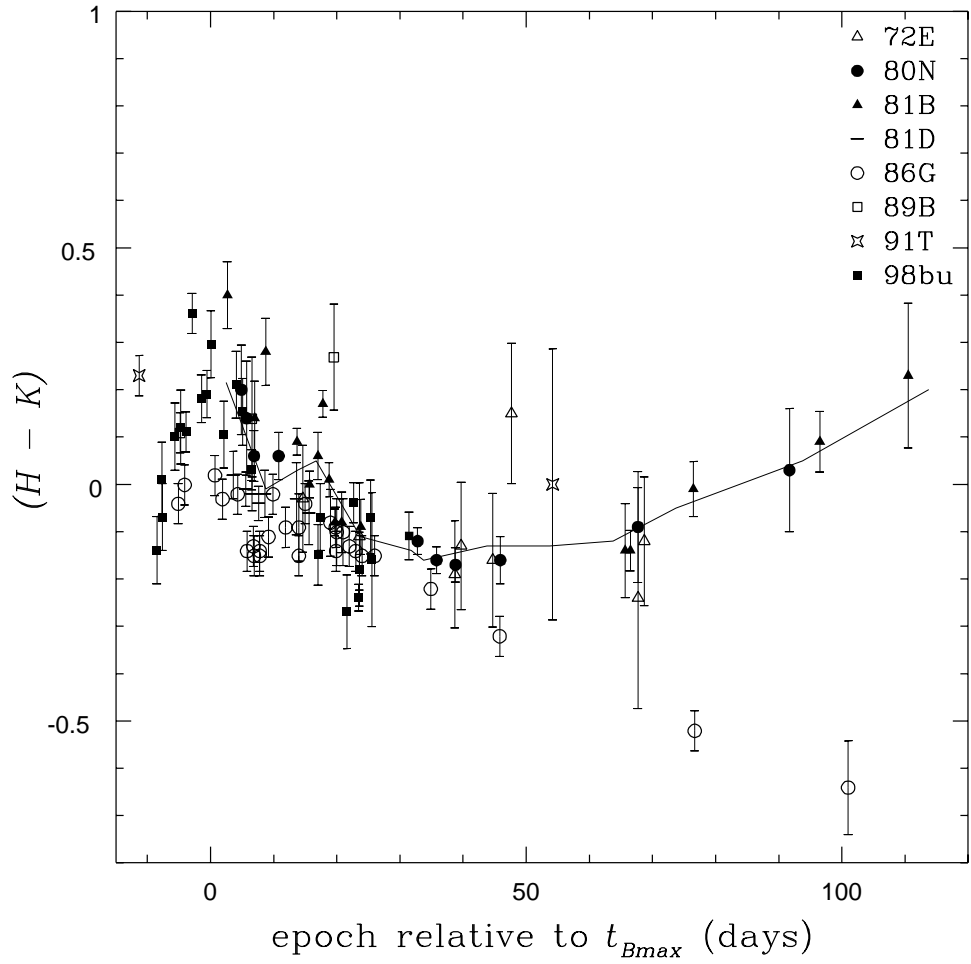}
\caption[]{Evolution of the $(H-K)$ colour of type~Ia supernovae.  Other
details as in Fig.~7 caption.}
\end{figure*}

\begin{figure*}
\vspace{14.0cm} \includegraphics{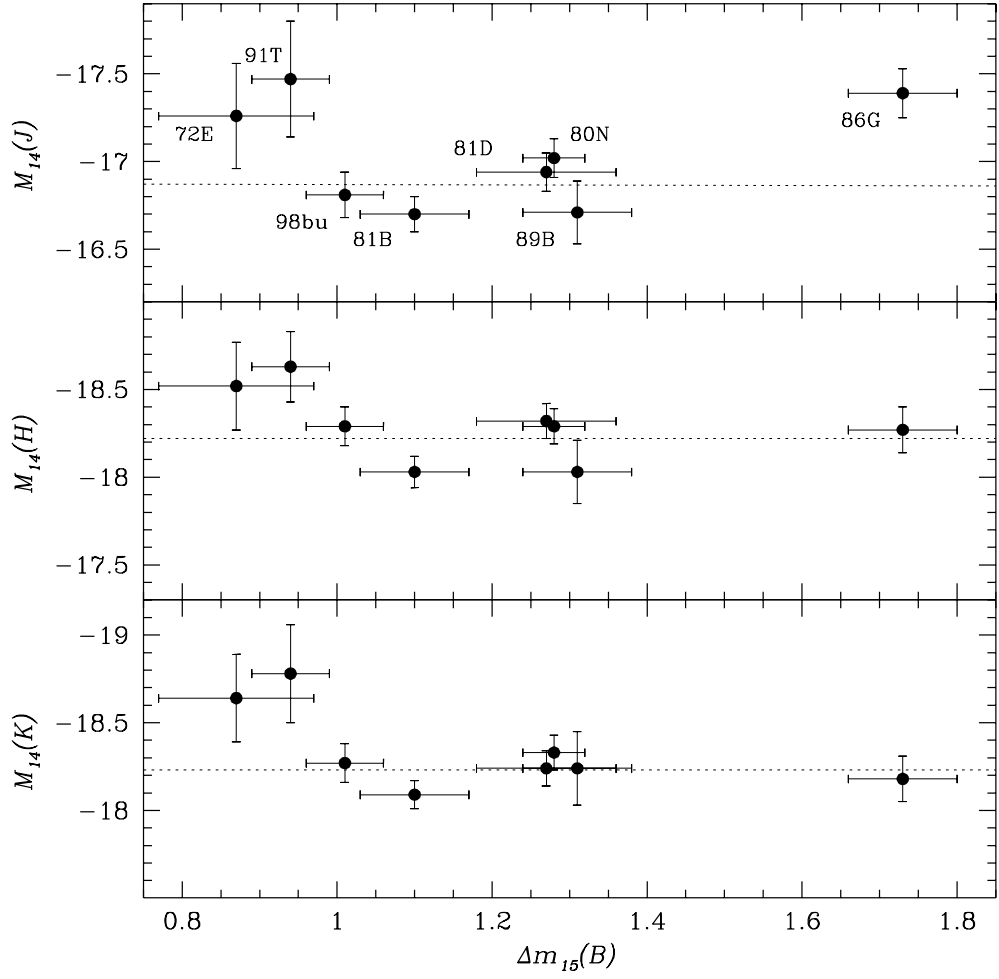}
\caption[]{Absolute $JHK$ magnitudes of type~Ia supernovae at
+13.75~days ($M_{14}$) plotted against the blue light curve decline
rate parameter $\Delta m_{15}(B)$.  The weighted mean values of
$M_{14}(J)$, $M_{14}(H)$ and $M_{14}(K)$, for the six ``IR-normal''
supernovae are indicated by the horizontal dotted lines.}
\end{figure*}

\end{document}